%% file: Text.tex
\documentclass[11pt]{article}

\usepackage[latin1]{inputenc}\usepackage{epsfig}
\usepackage{amsfonts}\usepackage{cite}

\title{\huge Real Functions for Physics }

\author{
  \Large Jorge L. deLyra \\
  Department of Mathematical Physics \\
  Physics Institute \\
  University of São Paulo }

\date{July 6, 2015}

\input{math-defs.tex}

\begin{document}\maketitle

\begin{abstract}
  \noindent
  A new classification of real functions and other related real objects
  defined within a compact interval is proposed. The scope of the
  classification includes normal real functions and distributions in the
  sense of Schwartz, referred to jointly as ``generalized functions''.
  This classification is defined in terms of the behavior of these
  generalized functions under the action of a linear low pass-filter,
  which can be understood as an integral operator acting in the space of
  generalized functions. The classification criterion defines a class of
  generalized functions which we will name ``combed functions'', leaving
  out a complementary class of ``ragged functions''. While the
  classification as combed functions leaves out many pathological objects,
  it includes in the same footing such diverse objects as real analytic
  functions, the Dirac delta ``function'', and its derivatives of
  arbitrarily high orders, as well as many others in between these two
  extremes. We argue that the set of combed functions is sufficient for
  all the needs of physics, as tools for the description of nature. This
  includes the whole of classical physics and all the observable
  quantities in quantum mechanics and quantum field theory. The focusing
  of attention on this smaller set of generalized functions greatly
  simplifies the mathematical arguments needed to deal with them.
\end{abstract}

\section{Introduction}

In a recent series of
papers~\cite{FTotCPI,FTotCPII,FTotCPIII,FTotCPIV,FTotCPV} we established a
correspondence between, on the one hand, real functions and generalized
real functions or distributions, all defined within a compact interval,
and on the other hand, certain analytic functions defined within the open
unit disk of the complex plane~\cite{ChurComp}. This correspondence
involves in a central way the Fourier coefficients of the real functions,
as well as the issue of the representability of the generalized functions
by their sequences of Fourier coefficients~\cite{ChurFour}. The
generalized functions as defined in~\cite{FTotCPV} are to be understood
loosely in the spirit of the Schwartz theory of
distributions~\cite{DistTheory}. The new correspondence that was
established allows one to deal with a large set of generalized functions,
either singular or not, via their representation in terms of analytic
functions, and therefore through the use of very solidly established
analytic procedures.

In a related paper~\cite{LPFFSaPDE} we introduced a set of linear low-pass
filters as tools that can be used to deal efficiently with divergent or
poorly convergent Fourier series resulting from the resolution of boundary
value problems of partial differential equations. In one of the
papers~\cite{FTotCPIII} of the series mentioned above these filters were
integrated into the structure of the aforementioned correspondence between
real generalized functions and complex analytic functions. This was done
via the introduction of complex low-pass filters within the open unit disk
of the complex plane, acting on complex analytic functions, that reproduce
the action of the real low-pass filters on the real functions when one
takes the limit from within the open unit disk to the unit circle.

Here we will use these elements to show that the first-order linear
low-pass filter can be used to establish a useful classification of all
the generalized functions. This will separate the set of all generalized
functions that one can define on the unit circle into two disjoint
subsets. One of these we will call the set of {\em combed functions}, the
other we will call the set of {\em ragged functions}. Although most of the
more profoundly pathological generalized functions are included in the
second subset, the classification is not based simply on smoothness, since
the Dirac delta ``function'' and many other singular generalized
functions, as well as many singular normal functions, are in fact
classified as combed functions.

We will argue that the set of combed generalized functions is sufficient
for all the needs of physics, in the role of tools for the description of
the observable aspects of nature. It should be pointed out that, while one
does not need real function in the continuum to describe aspects of nature
that are intrinsically discrete, such as the spin of elementary particles,
real generalized functions in the continuum can be advantageously used to
describe the behavior of physical quantities that depend on variables that
vary almost continuously, such as spatial positions. They can also be used
as very good approximations to describe those physical systems that
involve an extremely large number of degrees of freedom, such as extended
material objects. The argument is that combed generalized functions
suffice for these roles. The universe of applicability includes the whole
of classical physics, as well as all the observable quantities that vary
almost continuously in quantum mechanics and quantum field theory.

\section{The Low-Pass Filters}

Consider a real function $f(\theta)$ defined within the periodic interval
$[-\pi,\pi]$, or equivalently on the unit circle. In this paper we assume
that all real functions to be discussed are Lebesgue-measurable
functions~\cite{RealAnalysis}. Let us recall that for Lebesgue-measurable
real functions defined within a compact domain the conditions of
integrability, absolute integrability and local integrability are all
equivalent to one another, as discussed in~\cite{FTotCPV}. The real
functions we are to deal with here may be integrable in the whole domain,
or they may be what we call {\em locally non-integrable}, as defined
in~\cite{FTotCPV}. This means that they are not integrable on the whole
domain, but are integrable in all closed sub-intervals of the domain that
do not contain any of the non-integrable singular points of the function,
of which we assume there is at most a finite number. Therefore the term
``locally non-integrable'' is to be understood as meaning ``locally
integrable almost everywhere''. An integrable singularity is one around
which the asymptotic integral of the function exists, while around a
non-integrable one the asymptotic integral does not exist, or diverges to
infinity.

\begin{figure}[ht]
  \centering
  \fbox{
    \epsfig{file=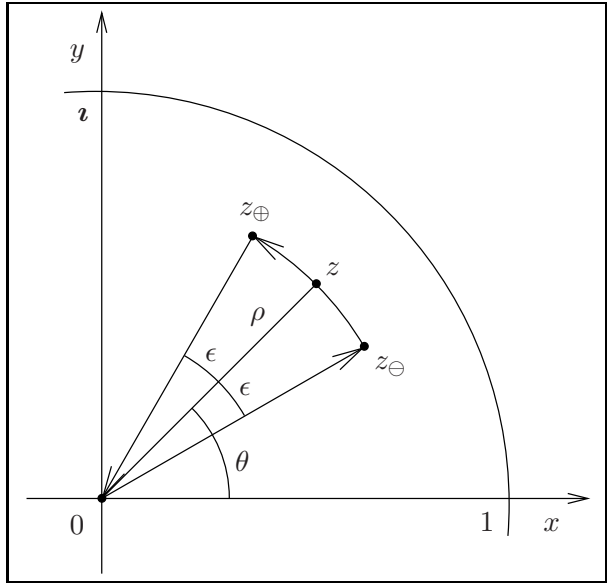,scale=1.0,angle=0}
  }
  \caption{Illustration of the definition of the complex first-order
    linear low-pass filter within the unit disk of the complex plane. The
    integral giving the average is taken over the arc of circle from
    $z_{\ominus}$ to $z_{\oplus}$. The points $z_{\ominus}$, $z_{\oplus}$
    and $0$ form a closed contour.}
  \label{Fig01}
\end{figure}

For such functions we may define the action of the first-order linear
low-pass filter as an operator acting on the space of real functions,
which from the real function $f(\theta)$ produces another real function
$f_{\epsilon}(\theta)$ by means of the integral

\begin{equation}\label{realfilt}
  f_{\epsilon}(\theta)
  =
  \frac{1}{2\epsilon}
  \int_{\theta-\epsilon}^{\theta+\epsilon}d\theta'\,
  f\!\left(\theta'\right),
\end{equation}

\noindent
which is well-defined at the point $\theta$ so long as the interval
$[\theta-\epsilon,\theta+\epsilon]$ does not contain any of the
non-integrable singularities of the original function. Since we are on the
unit circle, the parameter $\epsilon$ must satisfy $0<\epsilon\leq\pi$. In
this paper we will be interested mostly in the limit $\epsilon\to 0$, and
in this limit this definition suffices to determine $f_{\epsilon}(\theta)$
at all points except for the non-integrable singular points of
$f(\theta)$, that is to say almost everywhere, and strictly everywhere
within the domain of definition of $f(\theta)$ itself. The low-pass filter
can also be defined in terms of an integration kernel
$K_{\epsilon}\!\left(\theta-\theta'\right)$,

\begin{displaymath}
  f_{\epsilon}(\theta)
  =
  \int_{-\pi}^{\pi}d\theta'\,
  K_{\epsilon}\!\left(\theta-\theta'\right)
  f\!\left(\theta'\right),
\end{displaymath}

\noindent
where the kernel is defined as

\noindent
\begin{displaymath}
  % 
  % Redefining a length: space between rows.
  \renewcommand{\arraystretch}{2.0}
  \begin{array}{rclcc}
    K_{\epsilon}\!\left(\theta-\theta'\right)
    & = &
    \FFrac{1}{2\epsilon}
    &
    \mbox{for}
    &
    \left|\theta-\theta'\right|<\epsilon,
    \\
    K_{\epsilon}\!\left(\theta-\theta'\right)
    & = &
    0
    &
    \mbox{for}
    &
    \left|\theta-\theta'\right|>\epsilon.
  \end{array}
\end{displaymath}

\noindent
As was shown in~\cite{FTotCPIII}, this real operator acting on the real
functions can be obtained from a corresponding complex operator acting on
the inner analytic functions, in the limit from the open unit disk to the
unit circle, as follows. Consider an inner analytic function $w(z)$, with
$z=\rho\exp(\ii\theta)$ and $0\leq\rho\leq 1$. We define from it the
corresponding filtered complex function $w_{\epsilon}(z)$, using the real
angular range parameter $0<\epsilon\leq\pi$, by

\begin{equation}\label{compfilt1}
  w_{\epsilon}(z)
  =
  -\,
  \frac{\ii}{2\epsilon}
  \int_{z_{\ominus}}^{z_{\oplus}}dz'\,
  \frac{1}{z'}\,
  w\!\left(z'\right),
\end{equation}

\noindent
involving an integral over the arc of circle illustrated in
Figure~\ref{Fig01}, where the two extremes are given by

\noindent
\begin{eqnarray*}
  z_{\ominus}
  & = &
  z\e{-\ii\epsilon}
  \\
  & = &
  \rho\e{\ii(\theta-\epsilon)},
  \\
  z_{\oplus}
  & = &
  z\e{\ii\epsilon}
  \\
  & = &
  \rho\e{\ii(\theta+\epsilon)}.
\end{eqnarray*}

\noindent
This definition can be implemented at all the points of the open unit
disk. Note that if we make $z\to 0$, the integrand in
Equation~(\ref{compfilt1}) converges to a finite number, since the Taylor
series of $w(z)$ around $z=0$ has no constant term, given that $w(z)$ is
an inner analytic function. It follows that in that limit the integral
converges to zero, because the domain of integration becomes a single
point in the limit. Therefore we conclude that $w_{\epsilon}(0)=0$, which
means that the filter reduces to the local identity at $z=0$. Since on the
arc of circle we have that $z'=\rho\exp(\ii\theta')$ and hence that
$dz'=\ii z'd\theta'$, we may also write the definition of the complex
filtered function as

\begin{equation}\label{compfilt2}
  w_{\epsilon}(z)
  =
  \frac{1}{2\epsilon}
  \int_{\theta-\epsilon}^{\theta+\epsilon}d\theta'\,
  w\!\left(\rho,\theta'\right),
\end{equation}

\noindent
which makes it explicitly clear that what we have here is a simple
normalized integral over $\theta$. One can thus see that what we are doing
is to map the value of the function $w(z)$ at $z$ to the average of $w(z)$
over the symmetric arc of circle of angular span $2\epsilon$ around $z$,
with constant $\rho$. This defines a new complex function
$w_{\epsilon}(z)$ at that point.

Repeating what was done in~\cite{FTotCPIII} in the context of the
old-style inner analytic functions, and since it is a crucial part of our
present argument, let us now show that this complex function is in fact
analytic, and therefore that it is an inner analytic function according to
the newer definition given in~\cite{FTotCPV}, since we have already shown
that it has the property that $w_{\epsilon}(0)=0$. The definition in
Equation~(\ref{compfilt1}) has the general form of a logarithmic integral,
which is the inverse operation to the logarithmic derivative, as defined
and discussed in~\cite{FTotCPII}, where the logarithmic primitive of
$w(z)$ was defined as the integral

\begin{displaymath}
  w^{-1\!\ldot}(z)
  =
  \int_{0}^{z}dz'\,
  \frac{1}{z'}\,
  w\!\left(z'\right),
\end{displaymath}

\noindent
over any simple curve from $z'=0$ to $z'=z$ within the open unit disk, and
where we are using the notation for the logarithmic primitive introduced
in that paper. The logarithmic primitive $w^{-1\!\ldot}(z)$ is an analytic
function within the open unit disk, as shown in~\cite{FTotCPII}, and it
clearly has the property that $w^{-1\!\ldot}(0)=0$, so that it is an inner
analytic function as well. In order to demonstrate the analyticity of
$w_{\epsilon}(z)$ we consider the integral over the closed
positively-oriented circuit shown in Figure~\ref{Fig01}, from which it
follows that we have

\begin{displaymath}
  \int_{0}^{z_{\ominus}}dz'\,
  \frac{1}{z'}\,
  w\!\left(z'\right)
  +
  \int_{z_{\ominus}}^{z_{\oplus}}dz'\,
  \frac{1}{z'}\,
  w\!\left(z'\right)
  +
  \int_{z_{\oplus}}^{0}dz'\,
  \frac{1}{z'}\,
  w\!\left(z'\right)
  =
  0,
\end{displaymath}

\noindent
due to the Cauchy-Goursat theorem, since the contour is closed and the
integrand is analytic on it and within it. It follows that we have

\begin{displaymath}
  \int_{z_{\ominus}}^{z_{\oplus}}dz'\,
  \frac{1}{z'}\,
  w\!\left(z'\right)
  =
  w^{-1\!\ldot}(z_{\oplus})
  -
  w^{-1\!\ldot}(z_{\ominus}).
\end{displaymath}

\noindent
Since the logarithmic primitive $w^{-1\!\ldot}(z)$ is an analytic function
within the open unit disk, and since the functions $z_{\ominus}(z)$ and
$z_{\oplus}(z)$ are also analytic functions in that domain, it follows
that the right-hand side of this equation is an analytic function of $z$
within the open unit disk. We have therefore for the filtered complex
function

\begin{equation}\label{compfilt3}
  w_{\epsilon}(z)
  =
  -\,
  \frac{\ii}{2\epsilon}
  \left[
    w^{-1\!\ldot}(z_{\oplus})
    -
    w^{-1\!\ldot}(z_{\ominus})
  \right],
\end{equation}

\noindent
which shows that $w_{\epsilon}(z)$ is an analytic function as well. Since
we have already shown that $w_{\epsilon}(0)=0$, it follows that this
complex filtered function is an inner analytic function.

Let us now show that this complex low-pass filter reduces to the real
low-pass filter on the unit circle. If we write both the original function
$w(z)$ and the filtered function $w_{\epsilon}(z)$ in terms of their real
and imaginary parts, the expression in Equation~(\ref{compfilt2}) becomes

\begin{displaymath}
  \left[
    f_{\epsilon}(\rho,\theta)
    +
    \ii
    \bar{f}_{\epsilon}(\rho,\theta)
  \right]
  =
  \frac{1}{2\epsilon}
  \int_{\theta-\epsilon}^{\theta+\epsilon}d\theta'\,
  \left[
    f\!\left(\rho,\theta'\right)
    +
    \ii
    \bar{f}\!\left(\rho,\theta'\right)
  \right],
\end{displaymath}

\noindent
where $\bar{f}_{\epsilon}(\rho,\theta)$ and $\bar{f}(\rho,\theta)$ are the
harmonic conjugate functions respectively of $f_{\epsilon}(\rho,\theta)$
and $f(\rho,\theta)$. If we now take the $\rho\to 1$ limit to the unit
circle, we get from the real and imaginary parts of this expression

\noindent
\begin{eqnarray*}
  f_{\epsilon}(1,\theta)
  & = &
  \frac{1}{2\epsilon}
  \int_{\theta-\epsilon}^{\theta+\epsilon}d\theta'\,
  f\!\left(1,\theta'\right),
  \\
  \bar{f}_{\epsilon}(1,\theta)
  & = &
  \frac{1}{2\epsilon}
  \int_{\theta-\epsilon}^{\theta+\epsilon}d\theta'\,
  \bar{f}\!\left(1,\theta'\right),
\end{eqnarray*}

\noindent
so long as the integration interval on the unit circle does not contain
any non-integrable singularities of $w(z)$. Since the real part
$f_{\epsilon}(\rho,\theta)$ of $w_{\epsilon}(z)$ tends in the $\rho\to 1$
limit to the real generalized function $f_{\epsilon}(\theta)$
corresponding to the inner analytic function $w_{\epsilon}(z)$, we may
conclude that in the $\rho\to 1$ limit the complex filter reduces to the
definition of the real filter given in Equation~(\ref{realfilt}), for any
real generalized function $f(\theta)=f(1,\theta)$ that can be obtained as
the $\rho\to 1$ limit of the real part of an inner analytic function. The
same is true for the imaginary part, of course, which converges to the
corresponding ``Fourier Conjugate'' real generalized function, a concept
which was defined in~\cite{FTotCPI} and restated in~\cite{FTotCPV}.

\section{The {\boldmath $\epsilon\to 0$} Limit}

In a previously mentioned paper~\cite{LPFFSaPDE} it was shown that the
first-order linear low-pass filter, in its real form, tends to the
identity operator almost everywhere in the $\epsilon\to 0$ limit. In
Section~6 of Appendix~A of that paper one can find a simple proof that in
this limit the filtered function reproduces the value of the original
function whenever that function is continuous, that is

\begin{displaymath}
  \lim_{\epsilon\to 0}
  f_{\epsilon}(\theta)
  =
  f(\theta),
\end{displaymath}

\noindent
which holds at every point where $f(\theta)$ is continuous. At isolated
points of discontinuity where the two lateral limits of the original
function to the point of discontinuity exist, the filtered function
converges, in the $\epsilon\to 0$ limit, to the average of the two lateral
limits, as shown in Section~7 of that same Appendix,

\begin{displaymath}
  \lim_{\varepsilon\to 0}f_{\varepsilon}(\theta_{0})
  =
  \frac{1}{2}\left({\cal L}_{\oplus}+{\cal L}_{\ominus}\right),
\end{displaymath}

\noindent
where

\noindent
\begin{eqnarray*}
  {\cal L}_{\oplus}
  & = &
  \hspace{-1em}
  \lim_{\hspace{0.8em}\theta\to\theta_{0+}}f(\theta),
  \\
  {\cal L}_{\ominus}
  & = &
  \hspace{-1em}
  \lim_{\hspace{0.8em}\theta\to\theta_{0-}}f(\theta).
\end{eqnarray*}

\noindent
When the two limits coincide, and therefore the original function is
continuous at the point $\theta_{0}$, this of course reduces to the
previous property. At isolated points of non-differentiability where the
two lateral limits to that point of the derivative of the original
function exist, the derivative of the filtered function converges, in the
$\epsilon\to 0$ limit, to the average of these two lateral limits, as
shown in Section~8 of that same Appendix,

\begin{displaymath}
  \lim_{\varepsilon\to 0}
  \frac{df_{\varepsilon}}{d\theta}(\theta_{0})
  =
  \frac{1}{2}\left({\cal L'}_{\oplus}+{\cal L'}_{\ominus}\right),
\end{displaymath}

\noindent
where

\noindent
\begin{eqnarray*}
  {\cal L'}_{\oplus}
  & = &
  \hspace{-1em}
  \lim_{\hspace{0.8em}\theta\to\theta_{0+}}\frac{df}{d\theta}(\theta),
  \\
  {\cal L'}_{\ominus}
  & = &
  \hspace{-1em}
  \lim_{\hspace{0.8em}\theta\to\theta_{0-}}\frac{df}{d\theta}(\theta).
\end{eqnarray*}

\noindent
Of course, this implies that, at points where the function $f(\theta)$ is
differentiable, the derivative of the filtered function converges, in the
$\epsilon\to 0$ limit, to the derivative of the original function. From
all this we may conclude that, so long as there is only a finite number of
singular points, or at most an infinite but zero-measure set of such
points, in the $\epsilon\to 0$ limit the real filter becomes the identity
operator almost everywhere.

In addition to this, one can easily prove that the corresponding complex
filter always becomes {\em exactly} the identity operator in the
$\epsilon\to 0$ limit, within the open unit disk. This was commented on
in~\cite{FTotCPIII}, but since it is quite crucial to our current argument
let us repeat the demonstration here. We can see this from the
complex-plane definition in Equation~(\ref{compfilt3}). If we consider the
variation of $\theta$ between the extremes $z_{\oplus}$ and $z_{\ominus}$,
which is given in terms of the parameter $\epsilon$ by
$\delta\theta=2\epsilon$, and we take the $\epsilon\to 0$ limit of that
expression, we get

\noindent
\begin{eqnarray*}
  \lim_{\epsilon\to 0}
  w_{\epsilon}(z)
  & = &
  -\ii\,
  \lim_{\epsilon\to 0}
  \frac{w^{-1\!\ldot}(z_{\oplus})-w^{-1\!\ldot}(z_{\ominus})}{2\epsilon}
  \\
  & = &
  -\ii\,
  \lim_{\delta\theta\to 0}
  \frac{w^{-1\!\ldot}(z_{\oplus})-w^{-1\!\ldot}(z_{\ominus})}{\delta\theta}
  \\
  & = &
  z
  \lim_{\delta z\to 0}
  \frac{w^{-1\!\ldot}(z_{\oplus})-w^{-1\!\ldot}(z_{\ominus})}{\delta z},
\end{eqnarray*}

\noindent
where we used the fact that in the limit $\delta z=\ii z\delta\theta$.
Since we have that $\delta z=z_{\oplus}-z_{\ominus}$, we see that the
limit above defines the logarithmic derivative of $w^{-1\!\ldot}(z)$. In
addition to this, since that function is analytic in the open unit disk,
the limit necessarily exists. Therefore, we have

\noindent
\begin{eqnarray*}
  \lim_{\epsilon\to 0}
  w_{\epsilon}(z)
  & = &
  z\,
  \frac{d}{dz}w^{-1\!\ldot}(z)
  \\
  & = &
  w(z),
\end{eqnarray*}

\noindent
since we have here the logarithmic derivative of the logarithmic
primitive, and the operations of logarithmic differentiation and
logarithmic integration are the inverses of one another, as shown
in~\cite{FTotCPII}. We see, therefore, that this property within the open
unit disk is stronger than the corresponding property on the unit circle,
since in this case we have exactly the identity in all cases, while in the
real case we had only the identity almost everywhere. We have therefore
that, for {\em all} inner analytic functions $w(z)$,

\begin{displaymath}
  \lim_{\epsilon\to 0}w_{\epsilon}(z)
  =
  w(z),
\end{displaymath}

\noindent
which holds in the whole open unit disk. Since every inner analytic
function, filtered or not, corresponds to a real generalized function on
the unit circle in the $\rho\to 1$ limit, defined at all points where this
limit exists, the one-parameter family of inner analytic functions
$w_{\epsilon}(z)$ corresponds to a one-parameter family of real
generalized functions $f_{\epsilon}(\theta)$ on the unit circle. It is
therefore clear that as $w_{\epsilon}(z)$ approaches $w(z)$ in the
$\epsilon\to 0$ limit, so does $f_{\epsilon}(\theta)$ approach $f(\theta)$
in that same limit, at all points of the unit circle where the $\rho\to 1$
limit exists, that is, at least almost everywhere.

\section{The Classification}

The first-order low-pass filter can now be used to define a classification
of generalized functions. In order for the filter to be applicable, and
therefore for the classification to be feasible, the generalized functions
must be locally integrable almost everywhere, but they do not have to be
integrable on the whole domain. Observe, moreover, that we do {\em not}
have to assume that the generalized functions which we start with in this
argument were defined as $\rho\to 1$ limits of corresponding inner
analytic functions. Therefore, given a generalized function $f(\theta)$
which is locally integrable almost everywhere on the unit circle,
irrespective of whether or not it was defined as the $\rho\to 1$ limit of
an inner analytic function, we say that it is a {\em combed function} if
it is true that

\begin{equation}\label{combcond1}
  f(\theta)
  =
  \lim_{\epsilon\to 0}
  f_{\epsilon}(\theta),
\end{equation}

\noindent
where this recovery of the original function from the $\epsilon\to 0$
limit of the filtered function holds {\em everywhere}. Otherwise we say
that the generalized function is a {\em ragged function}. The results
obtained in~\cite{LPFFSaPDE} and discussed in the previous section now
imply that any continuous function is a combed function. We may consider
adopting a corresponding complex definition, using the corresponding
criterion for the inner analytic functions within the open unit disk.
Therefore we may classify an inner analytic function as combed if it
satisfies the condition that

\begin{equation}\label{combcond2}
  w(z)
  =
  \lim_{\epsilon\to 0}
  w_{\epsilon}(z),
\end{equation}

\noindent
where the recovery of the original function from the $\epsilon\to 0$ limit
of the filtered function holds everywhere within the open unit disk.
However, we showed in the previous section that this is the case for {\em
  all} inner analytic functions. This means that {\em every inner analytic
  function is combed within the open unit disk}.

Since the complex low-pass filter reduces to the real low-pass filter on
the unit circle, it follows at once that {\em all} the generalized
functions obtained as limits to the unit circle of the real parts of inner
analytic functions within the open unit disk are combed functions, which
we will refer to as ``combed generalized functions''. While this simply
reproduces the previously obtained results~\cite{LPFFSaPDE} for the case
of continuous real functions on the unit circle, it also means that the
Dirac delta ``function'' is a combed generalized function, which is quite
a remarkable fact. In this particular case it is not difficult to verify
this fact directly, but from the same result expressed by
Equation~(\ref{combcond2}) it follows that this is also true for all the
derivatives of the delta ``function'', of all orders, which is not so
simple and immediately apparent a fact.

Here is a simple direct proof that the delta ``function'' is a combed
generalized function. Consider a delta ``function''
$\delta(\theta-\theta_{0})$ centered at $\theta_{0}$. It is not difficult
to verify by direct calculation that the real filter with parameter
$\epsilon$, if applied to this generalized function, produces a
unit-integral rectangular pulse of width $2\epsilon$ and height
$1/(2\epsilon)$ centered at that same point. As one decreases $\epsilon$,
the $\epsilon\to 0$ limit of this one-parameter family of rectangular
pulses is one of the many ways commonly used as a definition of the delta
``function''. Therefore one recovers the delta ``function'' in the
$\epsilon\to 0$ limit, thus showing that the delta ``function'' is a
combed generalized function.

A similar argument can be constructed for the first derivative of the
delta ``function'', in which one must use some of the properties of the
delta ``function'' itself, and the fact that the derivative of the
integration kernel $K_{\epsilon}\!\left(\theta-\theta'\right)$ contains a
pair of delta ``functions''. However, in this case it is much simpler to
establish that the derivative of the delta ``function'' is a combed
generalized function by simply relying on the fact that it is the $\rho\to
1$ limit of an inner analytic function, as was shown in~\cite{FTotCPV}.
The same argument is valid, with equal ease, for all the derivatives of
the delta ``function'', of arbitrarily high orders.

It is also easy to see by direct calculation that any function with a few
isolated step-type discontinuities where the two lateral limits exist, and
which is otherwise continuous, is a combed function, so long as at the
points of discontinuity the function is defined as the average of the two
lateral limits. Note that this is the value to which the Fourier series of
the function converges, if that series is convergent at all. Once again,
this result can be easily derived from the fact that functions such as the
ones just described can be obtained as the $\rho\to 1$ limits of the real
parts of the corresponding inner analytic functions.

\subsection{Properties of Combed Functions}

Although the class of combed function includes many singular objects such
as the delta ``function'' and at least some locally non-integrable
functions, the generalized functions within the set have a significant
collection of common properties that considerably simplifies their
handling. Note that this is a large set of objects including many singular
ones of common use in physics, since combed function can be
non-differentiable, discontinuous, and may diverge to infinity at singular
points, which may or may not be integrable ones. For the purposes of this
section let us limit ourselves to the case in which the combed generalized
functions are integrable. This will make it easier for us to list their
main common properties, most of which have been demonstrated before. The
detailed extension of these properties to the complete set of combed
generalized functions is therefore postponed to some future opportunity.

\begin{itemize}

\item To start with, combed functions have no finite-spike singularities.
  In fact, any pathologies that do not have a definite non-zero effect on
  the integral of the function, such as finite discontinuities on a
  zero-measure subset of the domain, are simply eliminated by the
  application of the low-pass filter.

\item Since it is the $\epsilon\to 0$ limit of a filtered function, every
  combed function assumes at every given point the value given by the
  average of the two lateral limits of the original function to that
  point, so long as these limits exist,

  \begin{displaymath}
    \frac{{\cal L}_{\oplus}+{\cal L}_{\ominus}}{2},
  \end{displaymath}

  \noindent
  where

  \noindent
  \begin{eqnarray*}
    {\cal L}_{\oplus}
    & = &
    \hspace{-1em}
    \lim_{\hspace{0.8em}\theta\to\theta_{0+}}f(\theta),
    \\
    {\cal L}_{\ominus}
    & = &
    \hspace{-1em}
    \lim_{\hspace{0.8em}\theta\to\theta_{0-}}f(\theta),
  \end{eqnarray*}

  \noindent
  which in particular holds at all isolated discontinuities where the two
  lateral limits exist. Note that, since in the neighborhoods at the two
  sides of $\theta_{0}$, where $f(\theta)$ is continuous, the $\epsilon\to
  0$ limit of $f_{\epsilon}(\theta)$ reproduces $f(\theta)$, these two
  lateral limits of $f(\theta)$ coincide with the two corresponding
  lateral limits of $f_{\epsilon}(\theta)$ in the $\epsilon\to 0$ limit.

\item Every combed function $f(\theta)$ has a well-defined and unique
  sequence of Taylor-Fourier coefficients, from which it can be recovered
  strictly {\em everywhere} within its domain of definition. From the
  sequence of Taylor-Fourier coefficients one can always construct the
  corresponding inner analytic function $w(z)$. The combed real function
  is always given by the $\rho\to 1$ limit of the real part of its
  corresponding inner analytic function,

  \begin{displaymath}
    f(\theta)
    =
    \lim_{\rho\to 1}\Re[w(z)],
  \end{displaymath}

  \noindent
  everywhere within that domain, even if the Fourier series of that real
  function diverges everywhere.

\item If the Fourier series of a combed function converges at all, then it
  converges to that combed function everywhere in its domain of
  definition.

\item If the Fourier series of a combed function diverges, and so long as
  there is at most a finite number of dominant singularities of the
  corresponding inner analytic function on the unit circle, as they were
  defined in~\cite{FTotCPII}, it is possible to devise other expressions
  involving trigonometric series that converge to the real function, and
  that do so as fast as one may wish. We call these alternative
  trigonometric series ``center series'', and the algorithm to construct
  them is explained in~\cite{FTotCPII}.

\item Every combed function is the smoothest member of the class of
  zero-measure equivalent functions that it belongs to, as was discussed
  in Appendix~C of~\cite{FTotCPIV}. Given another member of that class,
  which is therefore a ragged function, there is no difficulty in
  ``combing'' it, that is in finding the combed function from which it
  differs only by a zero-measure function.

\item A ragged function can be combed by any one of the following methods:
  application to it of the low-pass filter in the $\epsilon\to 0$ limit;
  construction of its Fourier series, if convergent, and therefore of the
  limiting function of that series; construction of an expression
  containing an alternative trigonometric series, if the Fourier series
  fails to converge, and therefore of the limiting function of that
  alternative series; construction of its inner-analytic function and
  therefore of the limit of the real part of that complex function to the
  unit circle.

\end{itemize}

\noindent
In any given setting, if one limits the set of relevant real functions to
be discussed to combed generalized functions, then the use of the very
characteristic ``almost everywhere'' arguments in the harmonic analysis of
the functions become more sparse and better focused on the relevant
features of the functions. The possible exceptional points of the ``almost
everywhere'' arguments become only those singular points on the unit
circle that are brought about by the construction of the corresponding
inner analytic functions.

\section{Sufficiency for Physics}

In this section we argue, in a few different but related ways, that combed
generalized functions suffice to describe all physics. We mean this
statement to include all classical physics and all observables quantities
in quantum mechanics and quantum field theory. Note that we do not argue
that the use of combed generalized functions is necessary, only that it is
sufficient. There are a few ways in which one can argue this case, which
we will present in what follows. It should be noted in passing that the
limitation of the functions to the interval $[-\pi,\pi]$ is of no
consequence for this discussion of the physics applications, as shown in
Appendix~\ref{appredomint}.

\subsection{Fundamental Limitations on Precision}

Perhaps one of the most basic ways to argue the case starts from the
fundamental fact that all physical measurements of quantities that vary
almost continuously are necessarily made within finite and non-zero
errors. The representation of nature provided by physics is always an
approximate one. All physical measurements, as well as all theoretical
calculations related to them, of quantities which are represented by
continuous variables, can only be performed with finite amounts of
precision, that is, within finite and non-zero errors.

In fact, not only this is true in practice, both experimentally and
theoretically, but with the advent of relativistic quantum mechanics and
quantum field theory, it became a limitation in principle as well. This is
so because in its most fundamental aspect all particles in nature are
represented by fields, in the sense that the particles are just energetic
excitations of these fields. For finite particle energies the wavelengths
of these fields are finite and non-zero. Since it is not possible to
localize the particles with an accuracy that falls below the length scales
defined by those wavelengths, it follows that it is never possible to
measure the position of particles, or the dimensions of objects
constructed out of these particles, with mathematically perfect precision.

The use of real functions in physics is meant to be for the representation
of relations between physical variables that vary almost continuously. If
we assume that the physical quantities are observable, then they can at
best be measured or prepared (that is, set) within finite and non-zero
errors. A relation such as $y=f(x)$ where both $x$ and $y$ are observable
means that a finite-spike discontinuity in the function cannot have any
physical meaning. In order to detect such a spike in $y$ it would be
necessary to measure or prepare $x$ with infinite precision, which is
never possible, not only in practice but in principle. Therefore, there is
no possible physical meaning to ragged functions containing any such
finite-spike discontinuities.

\subsection{Representability in Momentum Space}

Another argument for the sufficiency of combed generalized functions for
the representation of physical quantities uses the representation of the
physics in momentum space. This representation is mathematically
equivalent to the use of the Fourier coefficients of the real generalized
functions as a way to represent those functions. The representation of the
physics in momentum space can always be constructed, and it is often found
to have a more direct and profound physical meaning than the original
representation in position space. This is so, in no small measure, because
the momentum-space modes that arise in this way represent distributed
quantities, and do not involve any attempt at localizing any objects with
mathematically perfect precision. Examples of this are everywhere, ranging
from normal modes of vibration in classical and quantum physics to the
construction of states of particles in quantum field theory.

The functions giving the solutions of physical problems are usually
solutions of boundary value problems of partial differential equations.
One of the most common and standard ways to solve such problems is via the
representation of the solutions in momentum space, which is equivalent to
the representation of the functions involved by their sequences of Fourier
coefficients. It follows at once that all solutions obtained in this way
are combed generalized functions. As was shown in the appendices
of~\cite{LPFFSaPDE}, the mere application of the low-pass filters will
usually improve rather than harm the representation of the physics by the
mathematics, even if the $\epsilon\to 0$ limit is not actually taken at
the end of the day. Moreover, when it is taken, since we are taking the
$\epsilon\to 0$ limit of a filtered function, we always end up with a
combed function.

Since two zero-measure equivalent generalized functions have exactly the
same set of Fourier coefficients, and thus cannot be distinguished from
each other by the use of those coefficients, it follows that if the
physics can be represented in momentum space at all, then the two
zero-measure equivalent generalized functions must represent exactly the
same physics. If we add to this the fact that the Fourier series, if
convergent at all, always converges to the combed function belonging to
that class of zero-measure equivalent generalized functions, then we must
conclude that the combed generalized functions are sufficient for the
representation of the physics involved. The same conclusion can be drawn
for every other way used to represent a generalized function by its
sequence of Fourier coefficients, such as through the corresponding inner
analytic function or through alternative expressions involving center
series, which are just other trigonometric series. This is so because all
such method of representation always produce combed functions.

\subsection{Representability by Differential Equations}

As a third argument for the sufficiency of combed generalized functions
for physics, we may go back to the fact that physics is often expressed by
second-order differential equations. Given an arbitrary second-order
differential equation to be satisfied by a real function $f(\theta)$, if
we are to consider the derivatives involved in the standard way, then the
function $f(\theta)$ must necessarily be twice-differentiable, and
therefore differentiable and continuous. Since every continuous function
is a combed function, this implies that $f(\theta)$ must be a combed
function, and in this case also a normal real function. However, just as
in the standard Schwartz theory of distributions, the introduction of
generalized functions represented by inner analytic functions allows us to
generalize the concept of differentiability and hence to enlarge the scope
of the differential equations.

If a real function $f(\theta)$ is not strictly differentiable at a given
singular point, we may still be able to uniquely attribute a value to its
derivative at that point, by the following sequence of operations: first
we go to the open unit disk of the complex plane and take the inner
analytic function $w(z)$ associated to that real function; second, since
the inner analytic function is always differentiable, we then take its
logarithmic derivative, which produces another inner analytic function
within the open unit disk; finally, we take the limit of the real part of
this new complex analytic function to the singular point on the unit
circle; it that limits exists (as it often will), we attribute that value
to the derivative of the original function at the singular point.

This can be further generalized, in case the limit to the unit circle does
not exist in a strict sense, by the global attribution of a singular
generalized function as the derivative of a strictly non-differentiable
function. In this way one may state, for example, that the derivative of a
function with a step-type singularity of height one at a certain point
contains a delta ``function'' centered at that point. From the point of
view of this more general definition of the derivative, all combed
generalized functions are differentiable, and in fact infinitely
differentiable, just as in the standard Schwartz theory of distributions.
We are therefore able to consider arbitrary generalized functions as
possible solutions of differential equations, so long as we reinterpret
these equations in terms of generalized derivatives.

Note that this extension of the scope of differential equations, if
executed via the representation of the generalized functions by inner
analytic functions, is done without leaving the realm of combed objects,
which in this case are combed generalized functions. Therefore, once again
we may argue that combed generalized functions suffice for the description
of physics.

\section{Conclusions}

We have shown that the real first-order low-pass filter, if taken in the
$\epsilon\to 0$ limit, can be used to classify all possible generalized
functions defined within a closed real interval into two disjoint classes,
a class of combed generalized functions and a complementary class of
ragged generalized functions. Although the first one contains, by and
large, more regular members and less singular members than the second, the
classification is not based simply on smoothness, since both classes do
contain singular members. In addition to this, given any ragged function,
there always is a corresponding combed function, from which it differs
only by a zero-measure function. In this way, the concept suggests itself
of the process of ``combing'' ragged functions. They can be ``combed'' so
long as they can be characterized by a sequence of Taylor-Fourier
coefficients, regardless of the convergence properties of the
corresponding Fourier series. In all cases there are several ways in which
this ``combing'' can be accomplished, including some that are
algorithmically sound and useful.

We have also shown that the class of combed functions contains all those
generalized functions that can be obtained as the $\rho\to 1$ limits of
inner analytic functions. This same class is also that which contains the
limiting generalized functions of all convergent Fourier series, as well
as the limiting generalized functions obtained from all convergent center
series built from the sequences of Fourier coefficients of real
generalized functions. At this time we are not able to ascertain whether
of not all combed generalized functions can be represented by inner
analytic functions. This is so because although we have shown
in~\cite{FTotCPV} that the set of inner analytic functions includes
representations of at least some locally non-integrable functions, we do
not yet know whether or not it contains all of them. In other words, it is
an open question whether or not all generalized functions which are
locally integrable almost everywhere are the derivatives of some order of
integrable generalized functions, and therefore can be represented by
inner analytic functions in the way presented in~\cite{FTotCPV}, using the
extended Taylor-Fourier coefficients introduced there.

It is interesting that, if one does not actually take the $\rho\to 1$
limit, but just gets sufficiently close to the unit circle to decrease
sufficiently the errors involved, then one can draw the conclusion that,
given any level of precision, there is an {\em analytic real function}
that represents the physics within that precision. If one considers the
restriction of the real part of any given inner analytic function to the
circle of radius $\rho=1-\delta$, for some arbitrarily small but non-zero
$\delta$, it becomes clear that this restriction is an analytic real
function, since it is infinitely differentiable everywhere. If the
limiting function represents some given physics within some error level
and there are no hard singularities (where the representation must break
down anyway), then the analytic real function obtained by this restriction
with a sufficiently small $\delta$ represents the same physics as the
function obtained in the $\rho\to 1$ limit, within the same error
level. We may thus conclude that, given any real function that represents
some physics within the finite and non-zero errors that are made
inevitable by the fundamental physical limitations on the determination of
all physical quantities that vary almost continuously, there is always a
real analytic function arbitrarily close to it, that represents the same
physics equally well, that is within the same level of precision.

The same conclusion can be reached if we start from the fact that the real
functions that carry physical meaning in physics applications can also be
represented by series in a basis of orthogonal functions, which is often
the case when the functions are solutions of boundary value problems
involving partial differential equations. Besides the Fourier basis of
functions, this includes such commonly used bases as those formed by
Legendre polynomials and those formed by Bessel functions of various
types, among many others. One property that all these bases share is that
they are infinite but discrete sets of analytic functions. If the series
representing the functions are convergent, then the real functions can be
represented by partial sums of the series within an arbitrarily given
finite level of precision. However, these partial sums are finite sums of
analytic functions and therefore are themselves analytic. In all these
cases it follows that the physics can always be represented by analytic
functions arbitrarily well, that is, within the finite and non-zero errors
that are mandated by the fundamental physical limitations on the
determination of all physical quantities that vary almost continuously.

As a final thought, one can pose the question of whether or not all
solutions to linear partial differential equations are necessarily combed
generalized functions. We believe that the right way to interpret this
question is to ask whether or not there is an even more general way to
extend the scope of differential equations that the transition from normal
functions to generalized functions, and from normal derivatives to
generalized derivatives, defined as we do here. So far as can be currently
ascertained, this seems to be an open question.

\appendix

\section{Appendix: Reduction of Domain Intervals}\label{appredomint}

The fact that we discuss the properties of real functions only within the
interval $[-\pi,\pi]$ is not a limitation from the physics point of view.
It is easy to show that all situations in physics application can be
reduced to this interval. Let us consider a physical variable $x$ within
the closed interval $[a,b]$, and let $g(x)$ be a real function describing
some physical quantity within that interval. The size of the interval does
not matter. For example, if $x$ is a measure of length, the interval could
be the size of a small resonant cavity or it could be the size of the
galaxy. In any case it is still a closed interval. The same is true if $x$
represents an energy, since it is always true that only limited amounts of
energy are available for any given experiment or realistic physical
situation. So long as $x\in[a,b]$, regardless of the magnitude or physical
nature of the dimensionfull physical variable $x$, we can rescale it to
fit within $[-\pi,\pi]$. It is a simple question of making a change of
variables, by defining the new dimensionless variable

\begin{displaymath}
  \theta
  =
  2\pi\,\frac{x-a}{b-a}-\pi,
\end{displaymath}

\noindent
so that $\theta\in[-\pi,\pi]$. The inverse transformation is given by

\begin{displaymath}
  x
  =
  \frac{b-a}{2\pi}\,\theta
  +
  \frac{b+a}{2}.
\end{displaymath}

\noindent
The function $g(x)$ can now be transformed into a function $f(\theta)$,

\begin{displaymath}
  g(x)
  =
  f(\theta),
\end{displaymath}

\noindent
where $f(\theta)$ describes the same physics as $g(x)$. Let us now
consider the transformation of the first-order low-pass filter from one
system of variables to the other. Given a value of $\theta$, if we vary it
by $\epsilon$, we have a corresponding variation of $x$, which we will
call $\varepsilon$. It then follows that we have a relation between
$\varepsilon$ and $\epsilon$,

\noindent
\begin{eqnarray*}
  x\pm\varepsilon
  & = &
  \frac{b-a}{2\pi}\,(\theta\pm\epsilon)
  +
  \frac{b+a}{2}
  \\
  & = &
  \left(
    \frac{b-a}{2\pi}\,\theta
    +
    \frac{b+a}{2}
  \right)
  \pm
  \frac{b-a}{2\pi}\,\epsilon
  \\
  & = &
  x
  \pm
  \frac{b-a}{2\pi}\,\epsilon
  \;\;\;\Rightarrow
  \\
  \varepsilon
  & = &
  \frac{b-a}{2\pi}\,\epsilon.
\end{eqnarray*}

\noindent
The limit $\epsilon\to 0$ is now seen to be clearly equivalent to the
limit $\varepsilon\to 0$, and the definition of the filtered function
$f_{\epsilon}(\theta)$ has a counterpart $g_{\varepsilon}(x)$,

\noindent
\begin{eqnarray*}
  f_{\epsilon}(\theta)
  & = &
  \frac{1}{2\epsilon}
  \int_{\theta-\epsilon}^{\theta+\epsilon}d\theta'\,
  f(\theta')
  \;\;\;\Rightarrow
  \\
  g_{\varepsilon}(x)
  & = &
  \frac{1}{2\varepsilon}
  \int_{x-\varepsilon}^{x+\varepsilon}dx'\,
  g(x').
\end{eqnarray*}

\noindent
It follows that if $f(\theta)$ is a combed function and we thus have that

\begin{displaymath}
  f(\theta)
  =
  \lim_{\epsilon\to 0}
  f_{\epsilon}(\theta),
\end{displaymath}

\noindent
then we also have that

\begin{displaymath}
  g(x)
  =
  \lim_{\varepsilon\to 0}
  g_{\varepsilon}(x),
\end{displaymath}

\noindent
so that $g(x)$ is also a similarly combed function. We may thus conclude
that it suffices to consider and discuss only the set of combed functions
within $[-\pi,\pi]$.

\end{document}

%% file: math-defs.tex
\newcommand{\ii}{\mbox{\boldmath$\imath$}}

\newcommand{\FFrac}[2]{{\displaystyle\frac{\displaystyle #1}{\displaystyle #2}}}

\newcommand{\e}[1]{\,{\rm e}^{#1}}

\newcommand{\ldot}{\mbox{\Large$\cdot$}\!}